# Challenges for RPCs and resistive micropattern detectors in the next few years

V. Peskov

*CERN, Geneva, Switzerland*

*and*

*UNAM, Mexico, Mexico*

*E-mail: vladimir.peskov@cern.ch*

**Abstract**

Nowadays RPCs are in a booming phase: they are successfully used in many experiments, including LHC; there are ambitious plans to use them in several upgrade detectors and in some new experiments as well as in various applications. The aim of this paper is to highlight the main challenges which the RPC community may face in the next few years and which were addressed in talks presented at this conference. Examples could be: new and difficult requirements from experiments (and their upgrades) and applications, optimization and improvements of the existing traditional detector designs, improvement of their characteristics (timing /rate performance, aging, dark current and so on), implementation of new more sensitive electronics, investigation of new materials, development of large- area detectors. We will also review the fast and very promising developments of another type of resistive electrode gaseous detector - micropattern detectors having at least one of their electrodes made of resistive materials. These innovative detectors combine in one design the best features of RPC (spark protection) and micropattern detectors (high granularity-high position resolution).

## I. Introduction

After a long and fruitful "wire detectors era" two major breakthroughs occurred in the field of gaseous detectors: developments of RPCs [1, 2] and invention of micropattern detectors (see for example review papers [3-5]). The previous ten RPC workshops were dedicated to the RPC developments, their validation, commissioning and extensive tests with beam particles or with cosmic rays. The developed RPC can be divided into two main categories: wide gap trigger RPCs [2] and narrow gap timing RPCs [6]. The trigger RPCs are made of bakelite or glass and have gaps in the range of 1-3 mm. Depending on the applied voltage, they can operate either in streamer or avalanche mode with a rate capability of the state of art designs ~300 Hz/cm$^2$ and 1-3 kHz/cm$^2$ respectively; their time resolution is about 1ns. The timing RPCs are made of glass, they have gaps typically below 0,3 mm and they are exploited in avalanche mode only achieving a time resolution of 50-70 ps. Both types of RPC can operate in multigap configurations introduced in [7]. Wide gap RPC- large area bakelite and glass - operating in streamer mode were already successfully used in several experiments, for example L3, OPERA, BaBar, BELLE, ARGO and others proving the maturity of this technology. The good performance of multigap timing RPCs was demonstrated by the HARP, STAR [8] and FOPI [9] collaborations.

The present 11$^{th}$ RPC Workshop is unique with respect to previous conferences because for the first time the operation of wide gap RPCs at LHC detectors (CMS, ATLAS and ALICE) were reported where they were used on an unprecedented scale (in total above 10$^4$ m$^2$) and high counting rate ~1kHz/cm$^2$(see for example talks [10-13]. Impressive results were also obtained with a large area (~150 m$^2$) ALICE TOF consisting of multigap timing RPCs [14]. During recent years multigap timing RPCs were also successfully operated in HADES experiment [15].

These results demonstrated that enormous efforts invested in the RPC technology and presented in the previous ten RPC conferences were not in vain and the RPCs are excellent detectors for muon trigger, tracking and time of flight measurements. Moreover, the resistive electrode approach is now successfully applied to micropattern gaseous detectors making them spark protected [16].

In the talks presented at this conference various authors identified and discussed the challenges which the RPC community will face in the next few years in high energy physics experiments, in astroparticle physics experiments and in applications. The aim of the paper is to summarize these challenges and identify possible ways of overcoming them. We will also review the fast and promising development of resistive electrode micropattern gaseous detectors. These spark-protected detectors are actually hybrids of RPCs and the "classical" micropattern detectors with metallic electrodes (GEM, MICROMEGAS, microstrip, microdot, micropin and so on).

## II. High energy physics experiments

## II.1. RPCs in running experiments

In spite of the great success of the RPCs in general, running experiments, of course, have a lot of challenges on a daily basis: stability in long term operation, temperature effect, dark current increase with time, dead channels, connectors, electronics, gas leaks and so on, just to mention a few. Relevant examples of how the RPC teams successfully cope with these daily challenges could be the ATLAS and CMS experiments. These teams introduced various sophisticated control systems to monitor in great detail RPC performance including low and high voltage power suppliers, trigger electronics, current and threshold monitoring, environmental sensors allowing a fast reaction on various effects in order to maintain high efficiency and stable operation of RPCs (see for example [17]). One of these effective measures includes HV working point corrections on changes of environmental pressure and temperature [18, 19] (see Fig.1).

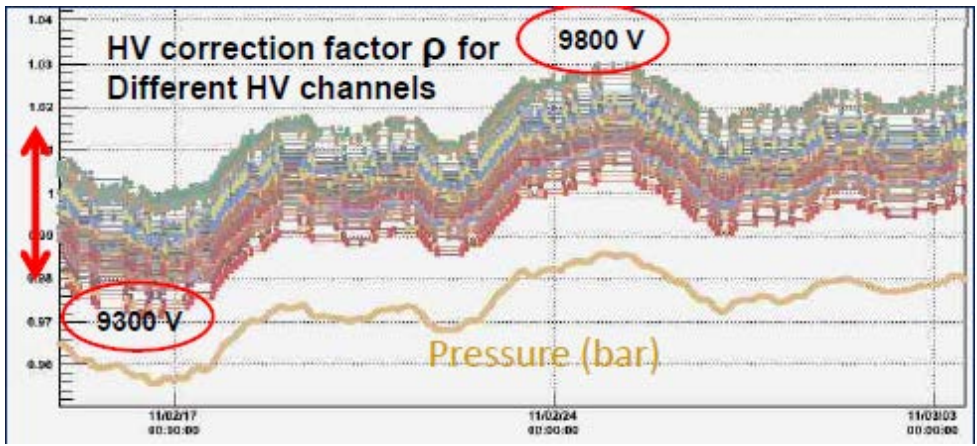

Fig.1. A schematic drawing illustrating RPCs HV working point corrections (see [18] for more details).

RPC current monitors allowed not only control of the RPC performance, but also an independent evaluation of the luminosity and the background in the cavern [20, 21] –see Fig. 2. A lot of effort was invested in gas purification, control of the gas quality and optimization of purifiers [22-24].
As a result the RPC systems at these experiments are running very reliably providing high quality data to the main data stream.
Certainly other running experiments also have various minor daily problems with their RPC (see, for example [8]), however all of them have so far been successfully solved proving that RPCs are promising detectors for future upgrades and new experiments.

### II.2. The next detector upgrades and new experiments
II.2.1. Trigger RPCs at LHC

The basic proposal for the LHC upgrade is, after seven years of operation, to increase the luminosity by up to a factor of 10, from the current nominal value of $10^{34}$ cm$^{-2}$ s$^{-1}$ to

$10^{35}$ cm$^{-2}$ s$^{-1}$. This means that the counting rate in the areas occupied by RPC will be increased accordingly. There will be two shut downs of the LHC in the coming years: in 2013 (so called phase I) and 2017 (phase II) during which the LHC itself and its detectors (CMS, ATLAS, and ALICE) will be upgraded to cope with new challenges.

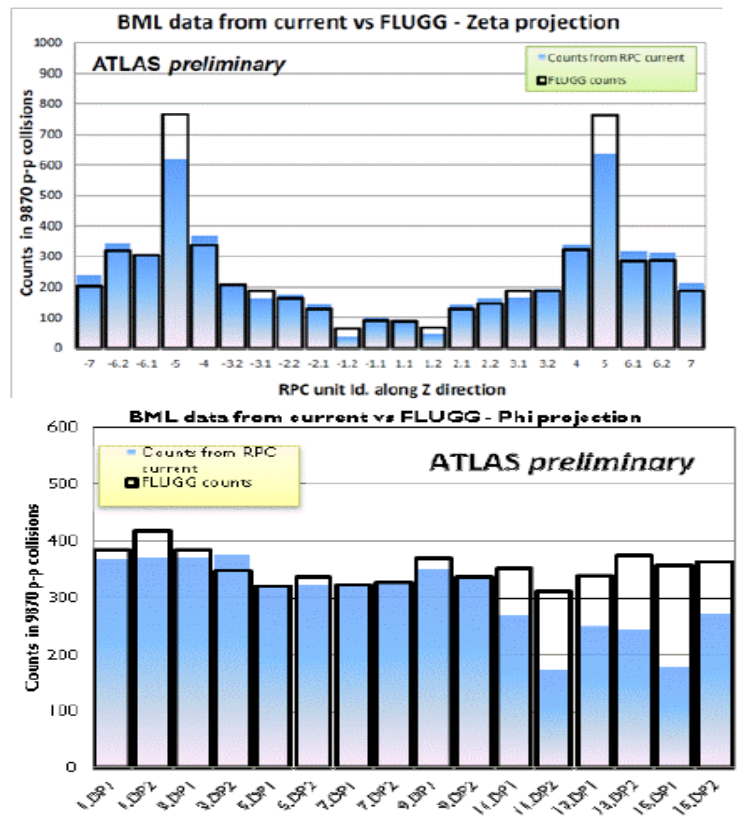

Fig.2 Comparison between the background measurements with RPC (upper figure) and predicted by a simulation -FLUGG- (lower figure) [20].

The CMS plan for 2013-2014 is to add detector's 4$^{th}$ endcup station (RE4), consisting of 144 « standard » bakelite RPCs in order to increase trigger efficiency in the region $\eta<1.6$ [25].This task requires a lot of international effort (trigger simulation, studies for the extended system, and details on the new HPL production) and their careful coordination: the chamber assembly and quality control procedures. Fig. 3 is shows schematically how the production and tests of these RPC will be spread over European and Asian scientific centers whereas their final assembly will be done at CERN. Obviously, this will not be an easy task although the CMS RPC team has great experience and is well- prepared for these challenges [26, 27].

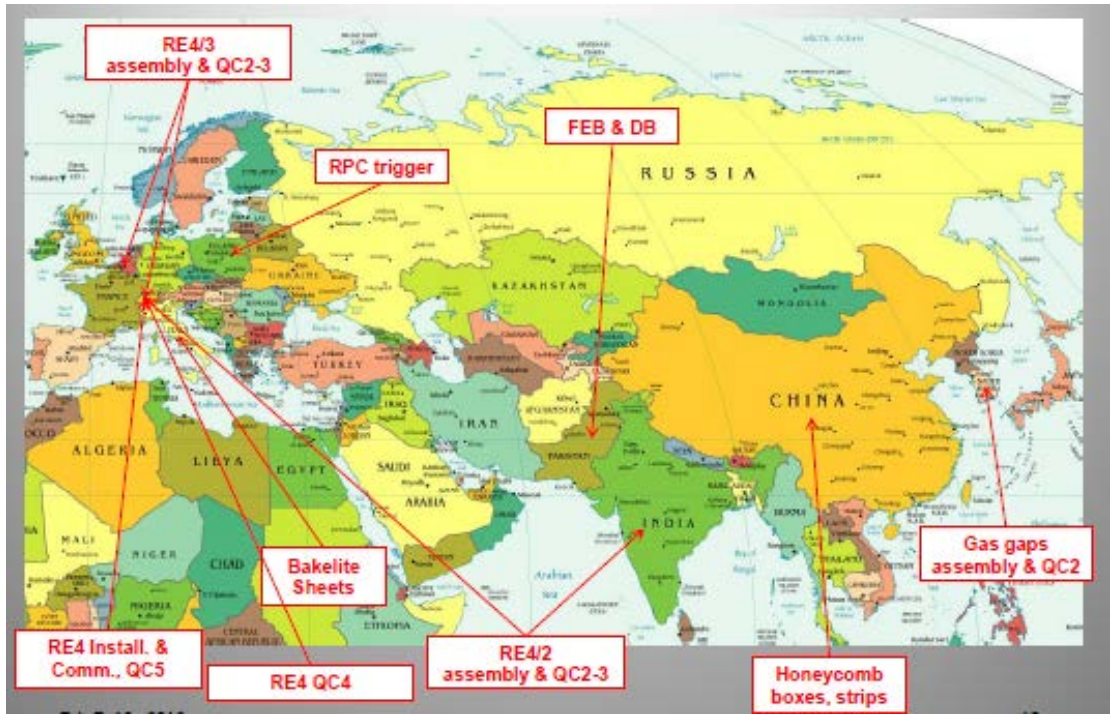

Fig. 3. A map showing locations of scientific centers involved in production, assembling and commissioning of RPCs for CMS [25].

There are even more challenges that the CMs RPC community will meet in preparation for phase II when, as shown in Fig. 4, the counting rate of RPC will be increased up to 10 kH/cm$^2$. The CMS RPC team started R&D work on developing high rate RPCs. Preliminary results are very promising indicating that RPCs made of low resistivity material can operate with high efficiency up to counting rates of 10 kHz/cm$^2$ [29]. A competing proposal is to use GEMs in the CMS forward region which offer very high rate capability ($10^6$Hz/mm$^2$) and excellent position resolution (better than 70μm)[28]. The final decision will be based on many considerations, for example time resolution (for the efficient background rejection a time resolution better than 1 ns is needed [30]), robustness/immunity to sparks, electronics, gas system and of course the cost.

ATLAS upgrade plans also focused on improving the trigger efficiency [31]. To do this a so-called small wheel is planned to be built and installed in 2017-see Fig.5. The small wheel detector should have a time resolution better than 0.5 ns, a rate capability of about 14 kHz/cm$^2$ and a spatial resolution of up to 0.3 mm. To cope with new challenges the ATLAS RPC team carried out extensive research and development programs with the aim to better understand and improve the RPC operation in avalanche mode [30-33]. The main directions for the RPC modifications are the following [30]: reduce the gas gap to 1mm, increase the sensitivity of the front end electronics and use multigap design (bi-gap considered so far). Additional measures may include the reduction of the electrode resistivity by up to $0.5\times10^{10}$Ωcm and also its thickness. Following these guidelines the detector layout schematically shown in Fig. 6 was proposed. It combines a tracking detector based on muon drift tubes (MDT) and a trigger detector using 1mm bi-gap RPCs.

| Forward Region | Rates Hz/cm² LHC ($10^{34}$ cm²/s) | High Luminosity LHC 2.3 x LHC | ($10^{35}$ cm²/s) Phase II |
|---|---|---|---|
| RB | 30 | Few 100 | kHz (tbc) |
| RE 1, 2, 3, 4  $\eta < 1.6$ | 30 | Few 100 | kHz (tbc) |
| Expected Charge in 10 years | 0.05 C/cm² | 0.15 C/cm² | ~ C/cm² |
| RE 1,2,3,4  $\eta > 1.6$ | 500Hz ~ kHz | Few kHz | Few 10s kHz |
| Total Expected Charge in 10 years | (0.05-1) C/cm² | few C/cm² | Several C/cm² |

Fig.4. Counting rate at which CMS RPC will operate with the increase of the LHC luminosity [28].

Currently a lot of effort is focused on attempts to operate 1mm gap RPC at reduced applied voltages in order to decrease the average charge in avalanches which will allow not only the rate characteristics of the RPCs to be improved, but also reduce their aging and the power consumption. This can be achieved with a high sensitive front-end electronics which is under development and being tested by the ALICE group [34]. As an alternative to the RPC approach the ATLAS team is also developing resistive MICROMEGAS (for more details see paragraph V).

ALICE muon trigger RPCs have electrodes of very low resistivity ($310^9$ Ωcm), however in order to be capable to operate at elevated rates in the future the ALICE RPC team is considering of operating their RPCs in avalanche mode instead of streamer mode as now[35]. This will require corresponding change of the electronics.

II.2.2. Multigap Timing RPCs in GSI upgrades and new experiments

In the next few years the new international accelerator facility FAIR, one of the largest research projects worldwide, will be erected at GSI. At FAIR an unprecedented variety of experiments will be possible. Thereby physicists from all around the world will be able to gain new insights into the structure of matter and the evolution of the universe from the Big Bang to the present.

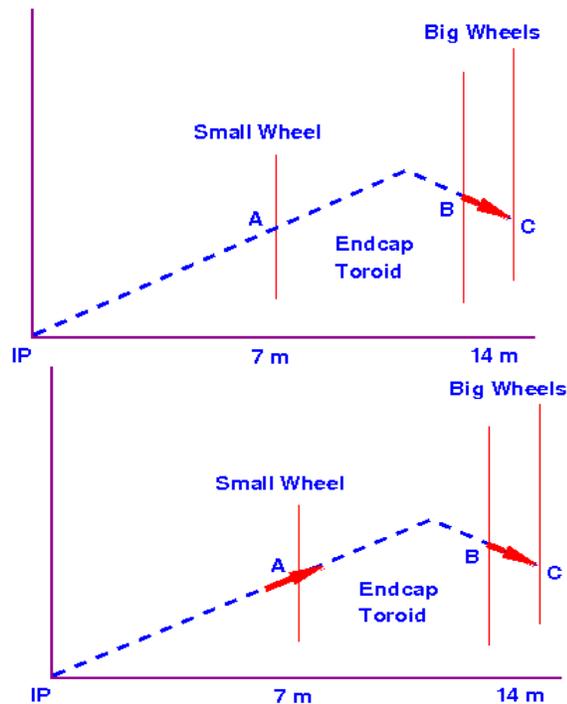

Fig. 5. The upper figure shows the current endcup trigger in which only a vector BC is measured. This leads to detection of fake tracks and also to bad momentum resolution. The small wheel (lower figure) provides an additional vector A; this will allow one to efficiently detect real track sand to improve Pt resolution [31].

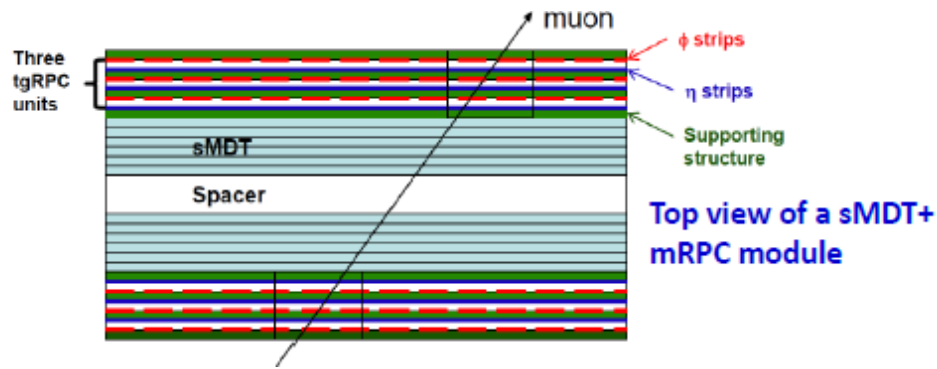

Fig. 6. Proposed detector module s for the ATLAS small wheel. It consists of a sandwich of MDTs and two RPC stations [31].

FAIR will provide antiproton and ion beams with unprecedented intensity and quality. In the final construction FAIR will consist of eight ring colliders of up to 1,100 meters in circumference, two linear accelerators and about 3.5 kilometers beam control tubes. The existing GSI accelerators serve as pre-accelerators.

a) CBM TOF

The CBM experiment is expected to be operational in 2017. Among other sub detectors, mostly traditional (vertex, silicon tracker, RICH and so on) the CBM collaboration is planning to build a large area (120m$^2$) TOF based on timing multistrip RPCs-see Fig.7 [36]. The real challenge here is that the timing RPCs in the central region of the TOF wall (marked in red color) should operate at a counting rate as high as 25 kHz/cm$^2$ with granularities in the range of 6-50 cm$^2$. At the same time the required intrinsic time resolution should be ~50 ps. The CBM TOF team is considering using RPCs made of glass with corresponding electrode resistivity in each zone marked in Fig. 7 in a differ color. For example, for the central region marked in Fig. 7 in red they are planning to use low resistivity $\rho \sim 10^{10} \sim \Omega$ "Chinese glass" (a kind of modification of the Pestov glass) [37] or ceramic which may have even lower resistivity[38].

b) R$^3$B experiment at FAIR

R3B is the name of the forthcoming upgrade of the actual LAND experimental setup at GSI. Narrow gap RPCs were proposed for building a time-of-flight wall for relativistic heavy ions (iTOF) with the main goal to achieve a time resolution of about 30 ps for the full iTOF (~50 ps for one individual RPC).The iTOF is dedicated detect ions after fragmentation reactions of heavy and medium mass nuclei at relativistic energies. The main challenge here is that the timing RPC performance with heavy ions is almost unknown and their timing properties could deteriorate with counting rate, even at rates >10Hz/cm$^2$ [39]. Another challenge is to use the minimum amount of materials in the RPC modules and the supporting structures[40].

II.2.3. STAR MTD

The STAR experiment is already equipped with a TOF based on a timing RPC [8]. The new challenge is to build a muon telescope detector (MTD) system. Presently several prototypes have already been tested showing promising preliminary results [41].

II.2.4. LEPS2/SPring-8

A new project to construct the second beam line for a laser-electron photon beam at Spring-8 (LEPS2) has started. LEPS2 will have a TOF system which should have rate capability > 1Hz/cm$^2$ and time resolution better than 50 ps. Multigap timing RPCs are well suited to these tasks and preliminary tests demonstrated this [42, 43].The challenge will be the current electronic improvements, developing RPCs with larger glass electrodes and larger pads and, of course, the installation and operation at LEPS2 in 2013.

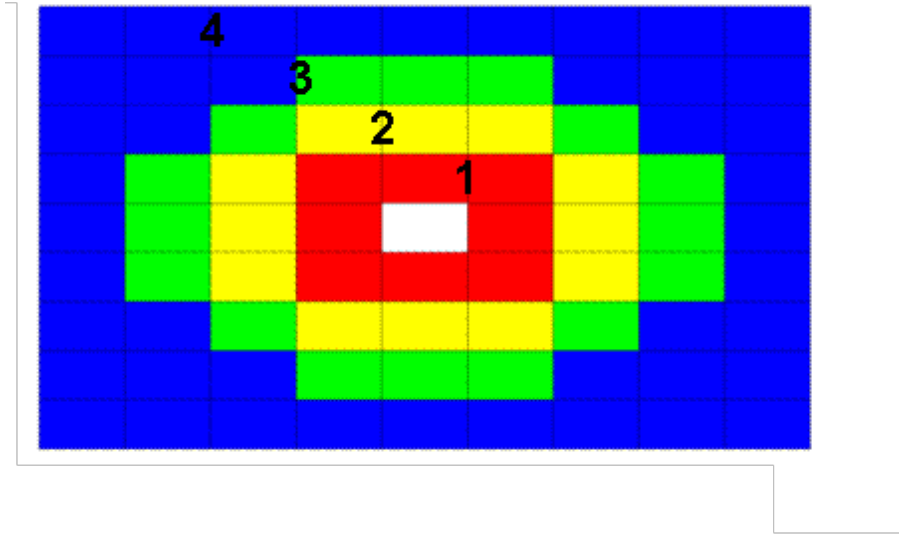

Fig.7. A schematic drawing of the CBM TOF wall showing regions (rectangular) occupied by timing RPCs with the following expected rate capabilities: red zone 8-25 kHz/cm$^2$, yellow zone 3.8-8 kHz/cm$^2$, green zone 1.5-3 kHz/cm2, blue zone 0.5-1.5 kHz/cm2 [36,37].

## III.   Astroparticle physics experiments

Previous conferences reported the technical design, problems, their solution and finally first results obtained by ARGO experiment using 2 mm gap bakelite RPCs covering an area of ~7000 m$^2$. The ARGO team successfully overcame a lot of technical challenges and now they enjoy the stage when exciting physical results are appearing [44]. Besides ARGO, OPERA [45] there are two new experiments using RPCs. One of them is the EEE project which is based on timing RPCs [46], another one is INO-ICAL (see Fig.8) which is planning to use large gap glass RPCs [47]. As any other experiment, these two collaborations are facing and will face a lot of challenges: gases and their distribution, possible leaks, implementation of the economic gas consumption, RPC long term stability, trigger schemes (ICAL) and so on (see [30] for more details).

## IV.   Additional challenges

The RPC community in general will have more difficulties and challenges than described above. Examples could be to find in the near future a replacement for Freon, $C_2H_2F_6$ and $SF_6$, a better understanding of the aging effect[48], development designs capable to operate at elevated rates [49], dark current effect, optimization of the gas circulation and purification, leak monitoring, development of low gas consumption schemes [50], developing simulation programs allowing RPC performance under various conditions to be predicted [51, 52], developing electronics, DAC, integration and much more.

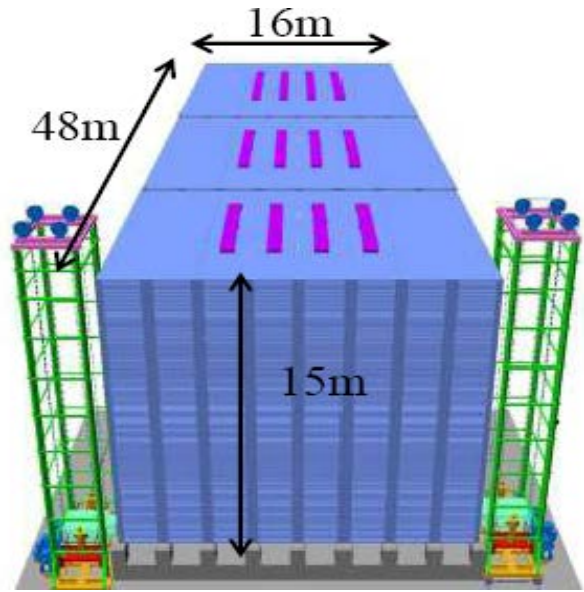

Fig. 8. A proposed design of the INI-ICAL- a 50 kt magnetized iron calorimeter. About 30,000 glass RPCs of about 2x2 m$^2$ in size are planning to be deployed as an active element of the ICAL [47].

Some challenges will come from new ideas (examples could be DHCAL using RPCs [53-55] and TOF tracker [56]) and from applications [57, 58]. However, based on previous experience one can be sure that the RPC teams are capable of solving all these problems.

There is an interesting tendency to merge RPC and micropattern detectors in one design [59- 61]. For example there are efforts from several groups of the RPC community to develop narrow gap(<1 mm) and high position resolution RPCc having a small pitch resistive or pad electrodes which somehow resemble micropattern detectors, but offering much better time resolution [30, 56].

Hence there is no doubt that RPC detectors will face a new momentum in developments accompanied with new challenges.

### V. RPC's new family members: micropattern gaseous detectors with resistive electrodes

Ordinary micropattern gaseous detectors (MSGC, MICOMEGAS, GEM and others) offer unprecedented 2D position resolution (20-40 μm) and thus are very attractive for many applications: tracking of charge particles, visualization of X-ray etc. [3-5].

However, as a consequence of their fine electrode structure these detectors are very fragile and for example, could be easily damaged by sparks. A few years ago, the first GEM-type micropattern detectors featuring resistive electrodes instead of metallic ones were developed [62]-see Fig.9.This detector was proposed for the ALICE RICH upgrade [63,64](see Fig. 10). The success of resistive GEMs triggered a series of similar developments, which are nowadays are being carried out by several other groups in the framework of the CERN RD51 collaboration [65]. As a result of these efforts it was demonstrated that the resistive electrode approach can be successfully applied to any micropattern gaseous detector and will make them robust and spark-protected [16, 66, 67]. The latest designs of resistive micropattern detectors have segmented (on strips) electrodes [68]. This approach turned out to be very fruitful allowing different designs of large-area spark-protected micropattern detectors oriented on various applications to be developed. For example, recently MICROMEGAS with resistive anode strips were developed for the ATLAS forward wheel upgrade [69]-see Fig. 11. Another design of resistive MICROMEGAS, so called Ingrid (see Fig. 12) was proposed as a gas tracker for the ATLAS upgrade [70].One can consider these detectors as a part of the RPC family. Indeed, metallic and resistive electrodes may be combined and still retain a property of the RPC: the total absence of violent discharges. The only requirement is that no gas gap will be delimited by two metallic electrodes.

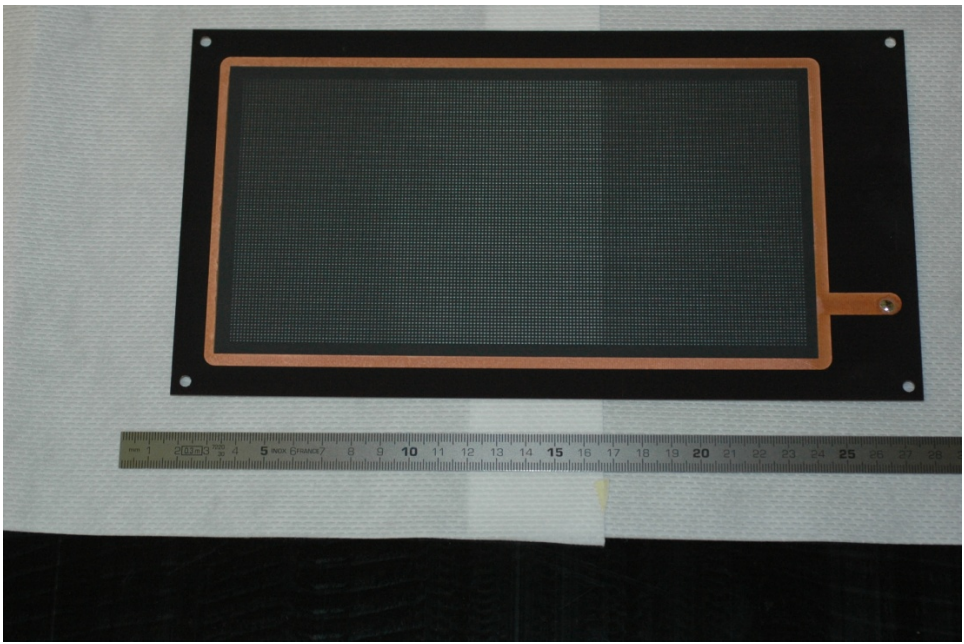

Fig.9. Photograph of GEM with resistive electrodes, developed in the framework of the ALICE RICH upgrade program [63].

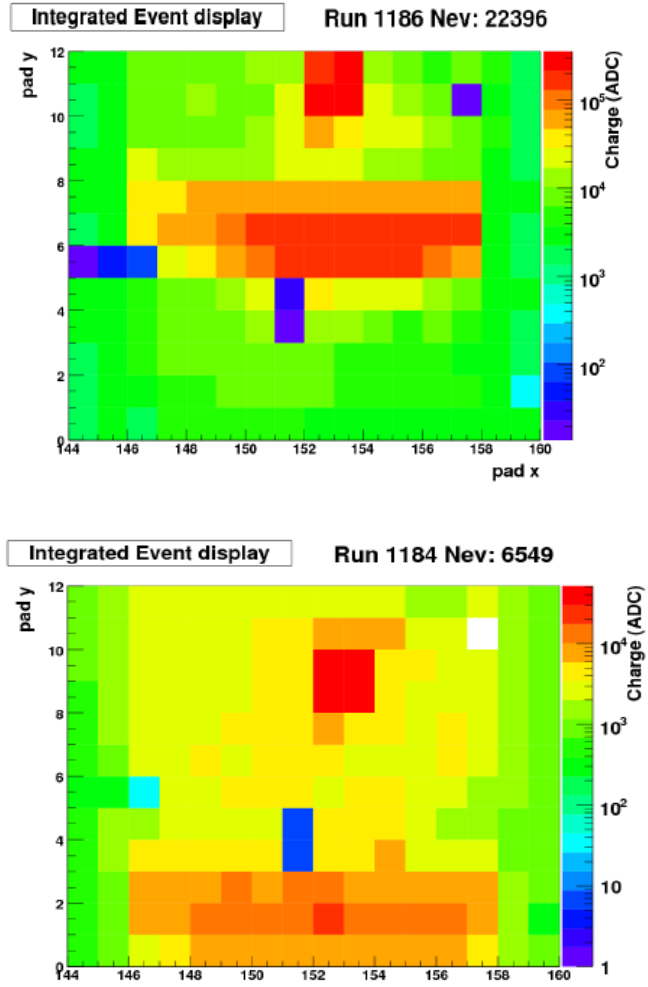

Fig. 10. Images of integrated events obtained during the beam tests of RICH prototype consisting of a $CaF_2$ radiator and a CsI coated resistive GEM oriented at 20° (upper figure) and 37° (lower figure) with respect to the pions. The spot at the top of each figure is the image of the particle beam and the horizontal band in the middle (upper figure) or in the bottom (lower figure) corresponds to the detected Cherenkov photons (see for example[64]).

### III. Conclusions

Nowadays RPCs are used or planned to be used in major high energy physics experiments, in some large scale astroparticle physics experiments and in various applications. Some challenges which the RPC community will face in the near future were identified in talks presented at this Conference. Let us conclude by summarizing them.

For the LHC RPC community: to develop RPC and electronics for Phase 1 upgrade and start preparations for Phase 2.

For TOF community: to achieve their ambition goals in the construction of high time and position resolution timing RPCs

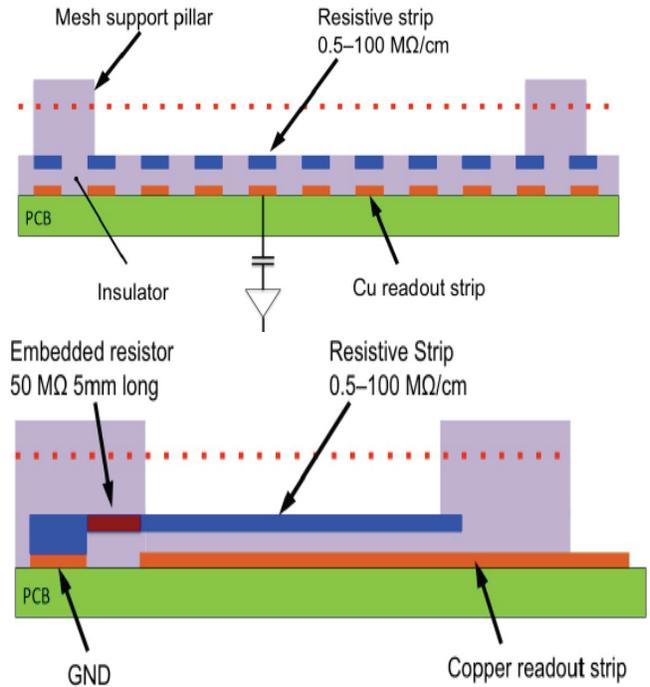

Fig. 11. Sketch of the resistive strip MICROMEGAS (two perpendicular cross sections) proposed for the ALICE upgrade [69].

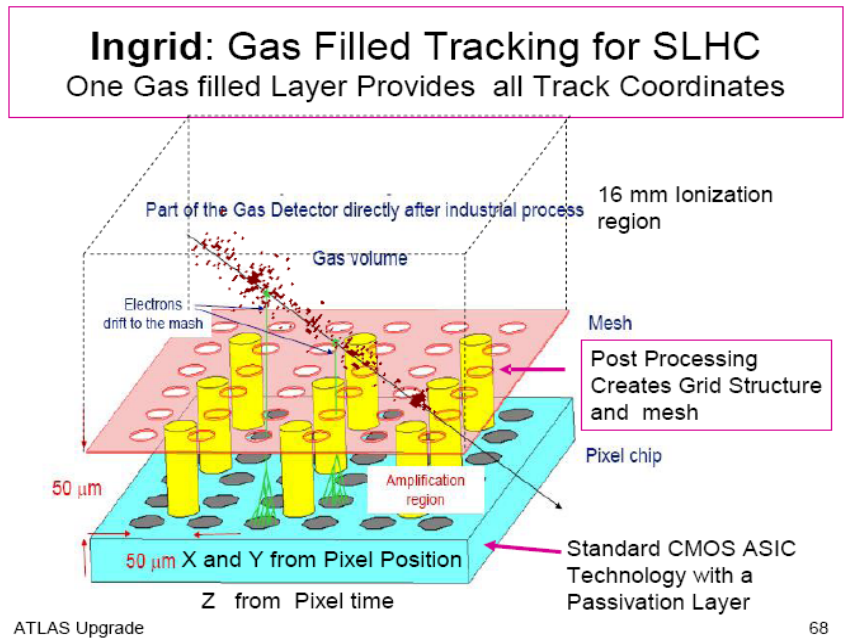

Fig 13. A schematic drawing of the Ingrid detector with resistive anode (5 layers of 1.4µm $Si_3N_4$), protecting CMOS ASIC from destruction by discharges [70].

Astrophysics community: construction of large area RPCs for IEE and INO-ICAL detectors, ensure their stable long-term operation.
"Application" community: commercialization of their devices (a very difficult task!)

For micropattern community: developing large area resistive detectors and investigating their timing properties (the ultimate goal will be to achieve a time resolution about 1ns), long-term stability and aging.

   Hence in the coming years the RPC and micropattern detector community will have a lot of exciting tasks, challenges and developments.

**References:**

[1]  V.V. Parhomchuk et al., *Nucl. Instrum. and Methods,* 93(1971) 269

[2]  B. Santonico eta l., *Nucl. Instrum. and Methods,* 187 (1981) 377

[3]  F. Sauli, *Nucl. Instrum. and Methods,* 477 (2002)1

[4]  T. Francke et al, Proc. Of the 42$^d$ Workshop of the INFN Eloisotron project " Innovative detectors for supecolliders", Erice, Italy, (2003) 158

[5]  V. Peskov , *"Progress in micropattern gaseous detectors and their applications", Proceedings of the 3$^d$ International workshop IWASI* 2009 (2009) 52

[6]  P. Fonte et al., *Nucl. Instrum. and Methods,* A443 (2000) 201

[7]  E. Cerron Zeballos et al., *Nucl. Instrum. and Methods,* A374(1996) 132

[8]  W.J. Llope, *Nucl. Instrum. and Methods,* A661, 2012 S110

[9]  M.Kis et al.,   Report at this Conference;
http://agenda.infn.it/conferenceTimeTable.py?confId=3950

[10] K. Bunkovski et al, Talk at this Conference;
http://agenda.infn.it/conferenceTimeTable.py?confId=3950

[11] F. Bossu et al., Talk at this Conference;
http://agenda.infn.it/conferenceTimeTable.py?confId=3950

[12] M. Gagliardi et al; Talk at this Conference;
http://agenda.infn.it/conferenceTimeTable.py?confId=3950

[13] P. Paolucci et al, Talk at this Conference;
http://agenda.infn.it/conferenceTimeTable.py?confId=3950

[14] A. Alici, et al., Talk at this Conference;
http://agenda.infn.it/conferenceTimeTable.py?confId=3950

[15] A. Blanco et al., Talk  at this Conference;
http://agenda.infn.it/conferenceTimeTable.py?confId=3950

[16] V. Peskov et al., *Nucl. Instrum. and Methods,* A661(2012) S153

[17] R. Berzano et al., Presentation at this Conference;
http://agenda.infn.it/conferenceTimeTable.py?confId=3950

[18] A. Polini et al., Report at this Conference;
http://agenda.infn.it/getFile.py/access?contribId=72&sessionId=2&resId=0&materialId=slides&confId=3950

[19] S. Constantini et al., Report at this conference;
http://agenda.infn.it/getFile.py/access?contribId=44&sessionId=2&resId=0&materialId=slides&confId=3950

[20] G. Aielli et al., Report at this Conference;
http://agenda.infn.it/contributionDisplay.py?contribId=73&sessionId=6&confId=3950

[21] M. Binini et al, Report at this Conference;
http://agenda.infn.it/getFile.py/access?contribId=74&sessionId=2&resId=0&materialId=slides&confId=3950

[22] S. Colafranceschi et al., Presentation at this Conference;
http://agenda.infn.it/conferenceTimeTable.py?confId=3950

[23] B. Mandelli et al., Talk at this Conference;
http://agenda.infn.it/conferenceTimeTable.py?confId=3950

[24] C. Lupi et al., Presentation at this Conference;
http://agenda.infn.it/conferenceTimeTable.py?confId=3950

[25] M. Tytgat et al., Presentation at this Conference;
http://agenda.infn.it/conferenceTimeTable.py?confId=3950

[26] S. Park, Talk at this Conference;
http://agenda.infn.it/conferenceTimeTable.py?confId=3950

[27] L.M Pant et al., Talk at this conference;
http://agenda.infn.it/conferenceTimeTable.py?confId=3950

[28] A. Scharma, Talk at this Conference;
http://agenda.infn.it/conferenceTimeTable.py?confId=3950

[29] I. Laktineh, private communication

[30] R. Santonico; http://agenda.infn.it/conferenceTimeTable.py?confId=3950

[31] J. Zhu et al., Talk at this Conference;
http://agenda.infn.it/conferenceTimeTable.py?confId=3950

[32] L. Han et al., Talk at this Conference;
http://agenda.infn.it/conferenceTimeTable.py?confId=3950

[33] L. Paolozzi et al, Talk at this Conference; http://agenda.infn.it/conferenceTimeTable.py?confId=3950

[34] R. Cardarelli et al., Talk at this Conference; http://agenda.infn.it/conferenceTimeTable.py?confId=3950

[35] M. Gagliardi, private communication

[36] I. Depner et al., Talk at this Conference; http://agenda.infn.it/conferenceTimeTable.py?confId=3950

[37] J. Wand et al., Talk at this Conference; http://agenda.infn.it/conferenceTimeTable.py?confId=3950

[38] A. Laso Garcia et al., Talk at this Conference; http://agenda.infn.it/conferenceTimeTable.py?confId=3950

[39] C. Paradela Dobarro et al., Talk at this Conference; http://agenda.infn.it/conferenceTimeTable.py?confId=3950

[40] E. Casarejos et al., Talk at this Conference; http://agenda.infn.it/conferenceTimeTable.py?confId=3950

[41] H. Chen et al., Talk at this Conference; http://agenda.infn.it/conferenceTimeTable.py?confId=3950

[42] N. Tomida et al., Talk at this Conference; http://agenda.infn.it/conferenceTimeTable.py?confId=3950

[43] C.Y. Hsien et al., Talk at this Conference; http://agenda.infn.it/conferenceTimeTable.py?confId=3950

[44] R. Luppa et al., Talk at this Conference; http://agenda.infn.it/conferenceTimeTable.py?confId=3950

[45] A. Paolini et al., Talk at this Conference; http://agenda.infn.it/conferenceTimeTable.py?confId=3950

[46] M. Abbrescia et al., Talk at this Conference; http://agenda.infn.it/conferenceTimeTable.py?confId=3950

[47] S. Dasgupta et al., Talk at this Conference; http://agenda.infn.it/conferenceTimeTable.py?confId=3950

[48] W. Yi et al., Talk at this Conference; http://agenda.infn.it/conferenceTimeTable.py?confId=3950

[49] K.S. Lee et al., Talk at this Conference; http://agenda.infn.it/conferenceTimeTable.py?confId=3950

[50] L. Lopes et al., Presentation at this Conference;
http://agenda.infn.it/conferenceTimeTable.py?confId=3950

[51] P. Fonte, Talk at this Conference;
http://agenda.infn.it/conferenceTimeTable.py?confId=3950

[52] G. Gonzales-Diaz, Talk at this Conference;
http://agenda.infn.it/conferenceTimeTable.py?confId=3950

[53] L. Xia et al., Talk at his Conference;
http://agenda.infn.it/conferenceTimeTable.py?confId=3950

[54] J. Repons et al., Talk at his Conference;
http://agenda.infn.it/conferenceTimeTable.py?confId=3950

[55] I. Laktineh et al., Talk at his Conference;
http://agenda.infn.it/conferenceTimeTable.py?confId=3950

[56] P. Fonte presentation at the section New Ideas;
http://agenda.infn.it/conferenceTimeTable.py?confId=3950

[57] P. Baesso et al, Talk at his Conference;
http://agenda.infn.it/conferenceTimeTable.py?confId=3950

[58] B. Pavlov et al., Talk at his Conference;
http://agenda.infn.it/conferenceTimeTable.py?confId=3950

[59] M. Byszewski et al., Talk at his Conference;
http://agenda.infn.it/conferenceTimeTable.py?confId=3950

[60] V. Peskov et al., Talk at his Conference;
http://agenda.infn.it/conferenceTimeTable.py?confId=3950

[61] T. Francke at al., *Nucl. Instrum. and Methods,* A508 (2003) 83

[62] R. Oliveira et al., *Nucl. Instrum. and Methods,* 576 (2007) 362

[63] https://twiki.cern.ch/twiki/pub/Sandbox/VHMPIDLoI/vhmpidLOI_v18.pdf

[64] P. Martinengo et al., *Nucl. Instrum. and Methods,* A639 (2011) 126

[65] Abstract book of the 2$^d$ Intern. Conf. MPGD-2011, Kobe, Japan, 2011:
book:http://ppwww.phys.sci.kobe-u.ac.jp/~upic/mpgd2011/Abstracts.pdf

[66] V. Peskov et al, JINST 7 C01005 (2012)

[67] V. Peskov, Talk at this Conference;
http://agenda.infn.it/conferenceTimeTable.py?confId=3950

[68] A. Di Mauro et al, *IEEE Trans. Nucl. Sci.,* 56 (2009) 1550

[69] M. Byszewski et al., Talk at this conference;
 http://agenda.infn.it/conferenceTimeTable.py?confId=3950

[70] M. Newcomer;
 http://www.hep.upenn.edu/HEP_website_09/Talks/Seminars/talks/2008_newcomer.pdf